# High-Q terahertz metamaterial from superconducting niobium nitride films

C. H. Zhang[1,2], J. B. Wu[1], B. B. Jin[1]*, Z. M. Ji[1], L. Kang[1], W. W. Xu[1], J. Chen[1] and P. H. Wu[1]*

[1]Research Institute of Superconductor Electronics (RISE), School of Electronic Science and Engineering, Nanjing University, Nanjing 210093, China

[2]Institute of Laser Engineering, Osaka University, 2-6 Yamadaoka, Suita, Osaka 565-0871, Japan

M. Tonouchi[2]

[2]Institute of Laser Engineering, Osaka University, 2-6 Yamadaoka, Suita, Osaka 565-0871, Japan

*Correspondence authors bbjin@nju.edu.cn and phwu@nju.edu.cn

We present in this letter low ohmic-losses terahertz (THz) metamaterials made from low-temperature superconductor niobium nitride (NbN) films. The resonance properties are characterized by THz time-domain spectroscopy. The unloaded quality factor reaches as high as about 178 at 8 K with the resonance frequency at around 0.58 THz, which is about 24 times as many as gold metamaterials with the same structure. The unloaded quality factor also keeps high as the resonance frequency increases, which is about 90 at 1.02 THz that is close the gap frequency of NbN film. All these experimental observations are well understood in the framework of Bardeen-Cooper-Schrieffer theory and equivalent circuit model. Our work offers an efficient way to design and make high-performance THz electronic devices.

Metamaterials (MMs) consisting of artificial metallic structure elements have achieved exotic electromagnetic phenomena, such as artificial magnetism[1], negative refractive index [2], superfocusing [3-4] and extraordinary transmission[5], which are absent in natural materials. The physical origin of these phenomena is from the resonant nature of the metallic structure elements [1, 6]. Meanwhile, the strong metallic loss of the resonant element will balance the amplification of evanescent wave in MMs, and hinder the implementation of the exotic phenomena mentioned above [1, 7]. In the microwave region, this loss is low, and the exotic electromagnetic phenomena mentioned above can still be easily demonstrated. However, as the frequency is pushed higher towards to terahertz (THz), this loss will increase greatly and have a large negative impact on the realization of exotic electromagnetic phenomena. An urgent goal is therefore to find a kind of MMs that have lower loss at THz for the realization and its practical applications of such phenomena.

Recently, two techniques are proposed to reduce the ohmic loss or increase the quality factor $Q$ of the resonance. The first one was to cool the metallic elements to liquid nitrogen or helium temperatures [8]. A 14% at 77 K and 40% at 10 K increases in $Q$ of the resonance was observed in experiment and simulation, respectively [8]. The second one was to replace the normal metals by superconductors, which can yield lower ohmic loss, compared with the first case. Superconducting THz metamaterial made from Nb film has demonstrated its low loss behavior [9]. This material was once used to show negative refractive index in the microwave region [10]. Meanwhile, the superconducting THz MMs based on yttrium-barium-copper (YBCO) oxide film were also reported [11-13]. However, its surface resistance at 0.5 THz is larger than that of copper around 10 K[14]. Besides, theoretical studies have shown that the surface resistance of YBCO increases with frequency more rapidly that that of Cu does. Thus, we can conclude that at our working temperature at 8 K and THz, $R_s$ of Cu is smaller than YBCO, indicating the normal metal has an advantage over YBCO for fabricating high Q resonator at THz.

However, the low ohmic loss of THz Nb MM can only be maintained below 0.7 THz due to the limitation of gap frequency $f_g$, ($f_g=2\Delta_0/h$, where $\Delta_0$ is the energy gap at 0 K and $h$ is the Plank constant). Therefore, the superconducting film with a higher $f_g$ will be expected to work at higher frequency. In this paper, we present a low ohmic loss superconducting THz MM made from superconducting NbN film, which has higher $f_g$ (≈1.2 THz) than Nb [15]. We successfully demonstrated experimentally the unloaded quality factor ($Q_u$) of 178 at 0.58 THz, which is about 24 times as many as the one from gold with the same pattern at the same temperature. Even at 1.02 THz, which is higher than $f_g$ of Nb film, $Q_u$ of 90 is achieved. The obtained results can be well explained by the Bardeen-Cooper-Schrieffer (BCS) theory and equivalent circuit model [15-17]. Our work offers a novel path to design and make low-loss and temperature-tunable THz MMs.

Our approach uses electric-field-coupled inductor-capacitor (ELC) resonator structure [18]. The MMs are a square array of ELC resonators. The photo of the part of a sample is shown in Fig.1a. The 200 nm-thick NbN film was deposited using RF magnetron sputtering on 500 μm thick MgO substrate. Superconducting transition temperature ($T_c$) of the NbN film is 15.8 K. Then the MMs were patterned with standard photolithograph and reactive ion etching (RIE) method [19]. We studied three NbN MMs with different resonance frequency $f_r$ at around 0.58 THz, 0.81 THz and 1.02 THz, respectively, which were below $f_g$ of Nb, between $f_g$ of Nb and NbN, and close to $f_g$ of NbN, respectively. The geometry and the notations of dimensions of an individual ELC structure are shown in Fig.1b. The values of dimensions are listed in Table I. A contrastive sample was also fabricated by gold film with same thickness and pattern with Sample I. THz time-domain spectroscopy (THz TDS) incorporated with a continuous flow liquid helium cryostat is used to characterized the MMs over a temperature region from 8 K to 300 K. THz transmission spectra are measured under normal incidence, using a bare MgO substrate as the reference.

Our attention is focused on the fundamental mode resonance properties of the MMs. The THz power transmission spectra of Sample I are shown in Fig. 2, which exhibit LC resonance mode around 0.58 THz. At low temperatures, e.g. 8 K, the superconducting MM exhibits the strongest resonance, indicated by the sharp THz transmission dip at 0.58 THz with the minimal power transmission of 0.00025. The resonance strength decreases with the increasing temperature, indicated by the broadening and lifting of the resonance dip. As the temperature is close to $T_c$, the transmission spectrum experiences remarkable changes, and then it almost keeps constant as the temperature is higher than $T_c$. Other two samples also show the similar behavior as the temperature varies. Meanwhile, a gold MM with the same pattern is also characterized, whose transmission spectra at 8 K and 300 K are shown in the inset of Fig. 2. The resonance frequency is almost temperature independent. The power transmission minimum is about 0.05. So, we can say the change of resonance properties of gold MMs with temperature is imperceptible.

Fig. 3 shows the temperature dependence of $Q_u$ of the three NbN MMs. According to the Eq. (1) in Ref. 9, the coupling coefficient $\beta$ is *2 [P($\omega$)/P($\omega_0$)]$^{0.5}$-1,* where *P($\omega$)* and *P($\omega_0$)* represent the power transmission at the frequency far away from $\omega_0$ and at $\omega_0$, respectively. Then, we used this equation again to fit the measured resonance curves to obtain $Q_u$. At around 8 K, the $Q_u$ value of sample I can reach as high as 178. As the temperature increases, $Q_u$ decreases and keeps constant of about 2 above $T_c$. In contrast, $Q_u$ of the gold metamaterials is only about 7.6 and temperature independent. So, the superconducting NbN MMs has $Q_u$ of 24 times as many as the gold MMs. For sample II resonating at 0.81 THz, which is higher than $f_g$ of Nb film, $Q_u$ is about 120 at 8 K. Even resonating at 1.02 THz for sample III, which is close to $f_g$ of NbN $Q_u$ is about 90. This provides an efficient path to achieve high-$Q$ MMs even at high THz frequency, which is much better than Nb and YBCO.

Although our $Q_u$ is high at THz, the loaded quality factor $Q_l$ is just about 3, which is too low for practical applications. The physical reason is that the present structure is easy to couple with the free space, leading to a large coupling loss [20]. However, we believe that the coupling loss can be greatly reduced by optimizing the structure design such as using an asymmetric structure proposed by Fedotov. Having been able to get a $Q_u$, we are trying such new structures in the hope to get a higher $Q_l$.

The temperature dependence of transmission minimum is depicted in Fig.4. According to equivalent transmission model, it equals to [21]

$$T = \left| \frac{1+n_s}{1+n_s + Z_0/Z} \right|^2 \tag{1}$$

Where $Z_0$ is the impedance of vacuum, and $n_s$ is the refractive index of substrate. Here, $Z$ is the total impedance of one unit. For superconductor film with finite thickness of $d$, the effective surface impedance $Z_s$ could be calculated from the following formula [21]:

$$Z_S = (R_S + jX_S)\coth(d/\delta) = \sqrt{\frac{j\omega\mu_0}{\sigma}}\coth(d/\delta) \qquad (2)$$

where σ is the complex conductivity of NbN film, $\delta = X_s/\omega\mu_0$ is the penetration depth, and $\mu_0$ is the permeability of vacuum. $\sigma = \sigma_1 + j\sigma_2$ can be described in the framework of the BCS theory. Its temperature dependent can be calculated assuming the normal state conductivity, $\Delta(0)/k_BT_c$ and γ to be $1.2\times10^6$ S/m, 2.0 and 40 THz, respectively [15]. Here, $\Delta(0)$, $k_B$ and γ are energy gap at 0 K, Boltzmann constant and scattering rate respectively. A little change of the values in the normal state and $\Delta(0)/k_BT_c$ is to get a better agreement between the experiments and calculations. For each ELC unit, the reactance $X$ is zero at resonance, so the equivalent resistance $R \cong [(L_{loop}-g)/t]R_s$, where $L_{loop}$ is the equivalent length of current loop [19, 22]. As two current loops are shunt-wound, the total $Z$ is $R/2$ at resonance frequency.

The calculated temperature dependent transmission minimum is also plotted in Fig. 4, which agrees well with the experimental results. The discrepancy of transmission minimum at relative low temperature is due to the following physical reasons. The first one is that calculated values are outside the dynamic range of the THz spectroscopy. The second one is that residual surface resistance of the film, which do not allow for further decrease of $R_s$ at very low temperature, leading to a saturation of transmission minimum. The above agreement implies that coupling of MM to the free space could be described within the equivalent circuit model.

In conclusion, we successfully fabricated and demonstrated the low loss THz MM from superconducting NbN film. $Q_u$ can be about 24 times as many as the one of gold MMs at 0.58 THz. The high $Q_u$ performance of NbN MMs could remain well even at higher frequency of 1.02 THz which is close to the gap frequency of NbN. The THz resonance was analyzed theoretically using the BCS theory and RLC model. The calculated results agreed well with the experimental results. We hope that our results can contribute to solve the loss problem of THz MMs and development of THz devices based on superconducting film.

## Acknowledgments:

The work is supported by the National Basic Research Program of China under Grants No. 2011CBA00107 and No. 2007CB310404, NSFC program under contract No. 61071 009.

## Tables:

Table I: The parameters of the ELC unit cells and the resonant frequencies $f_r$ for three samples are listed in Table I. All units are in µm.

|  | g | w | t | l | p | $f_r$(THz) |
|---|---|---|---|---|---|---|
| Sample I | 5 | 10 | 5 | 50 | 60 | 0.58 |
| Sample II | 3.5 | 7 | 3.5 | 35 | 42 | 0.81 |
| Sample III | 2.5 | 5 | 2.5 | 25 | 30 | 1.02 |

# Figure captions

Fig. 1 (a) Scanning electron microscopy images of superconducting THz metamaterial, E and H represent the electric field and magnetic field. (b) Geometry and dimensions of an individual unit cell: orange and gray parts are thin film (NbN) and substrate (MgO), respectively.

Fig.2 Transmission spectra of the NbN metamaterial (Sample I) at various temperatures, the inset shows the transmission spectra of Au metamaterial at 8 K and 300 K.

Fig. 3 The unloaded quality factor ($Q_u$) of NbN MMs as a function of temperatures which has the resonance frequency of around 0.58 THz, 0.81 THz and 1.02 THz at 8 K, respectively. The lines are guided for clarity.

Fig. 4 Transmission minimums of the three NbN MMs as a function of the temperature. The line is the theoretical results and the symbols show the experimental results.

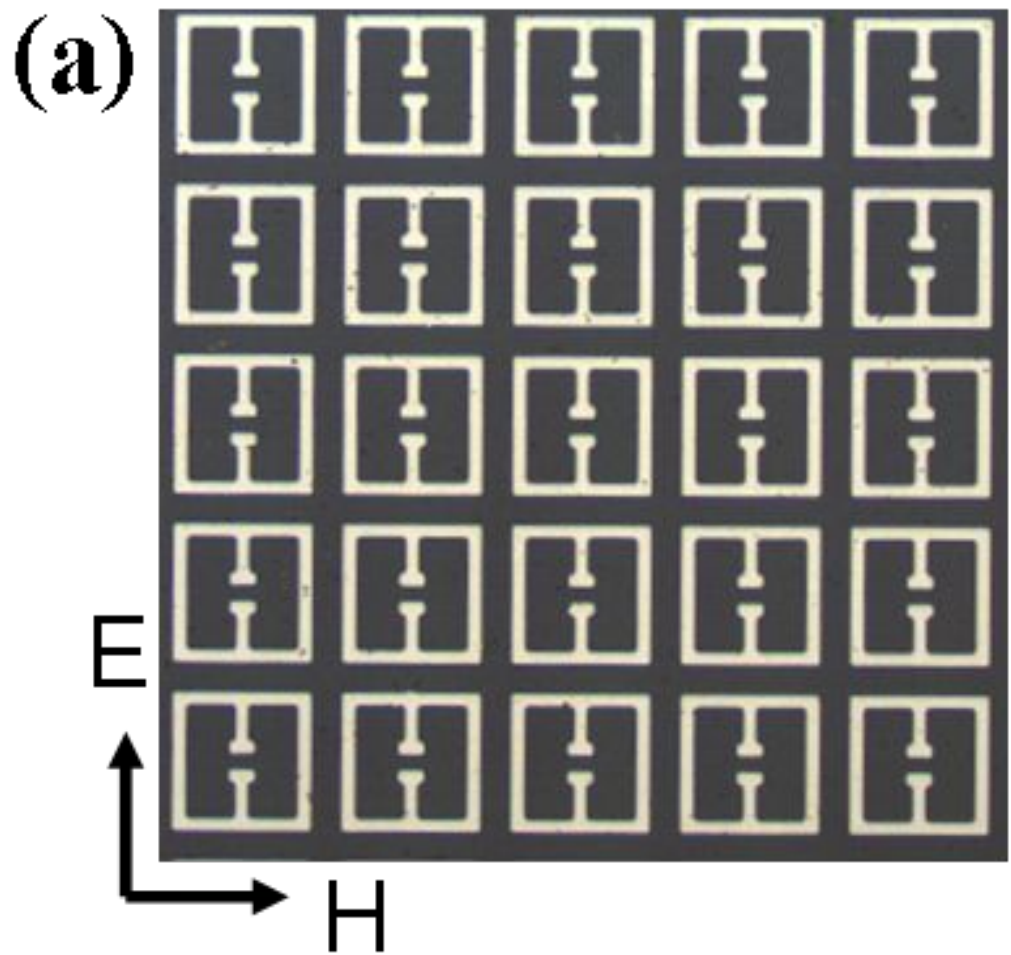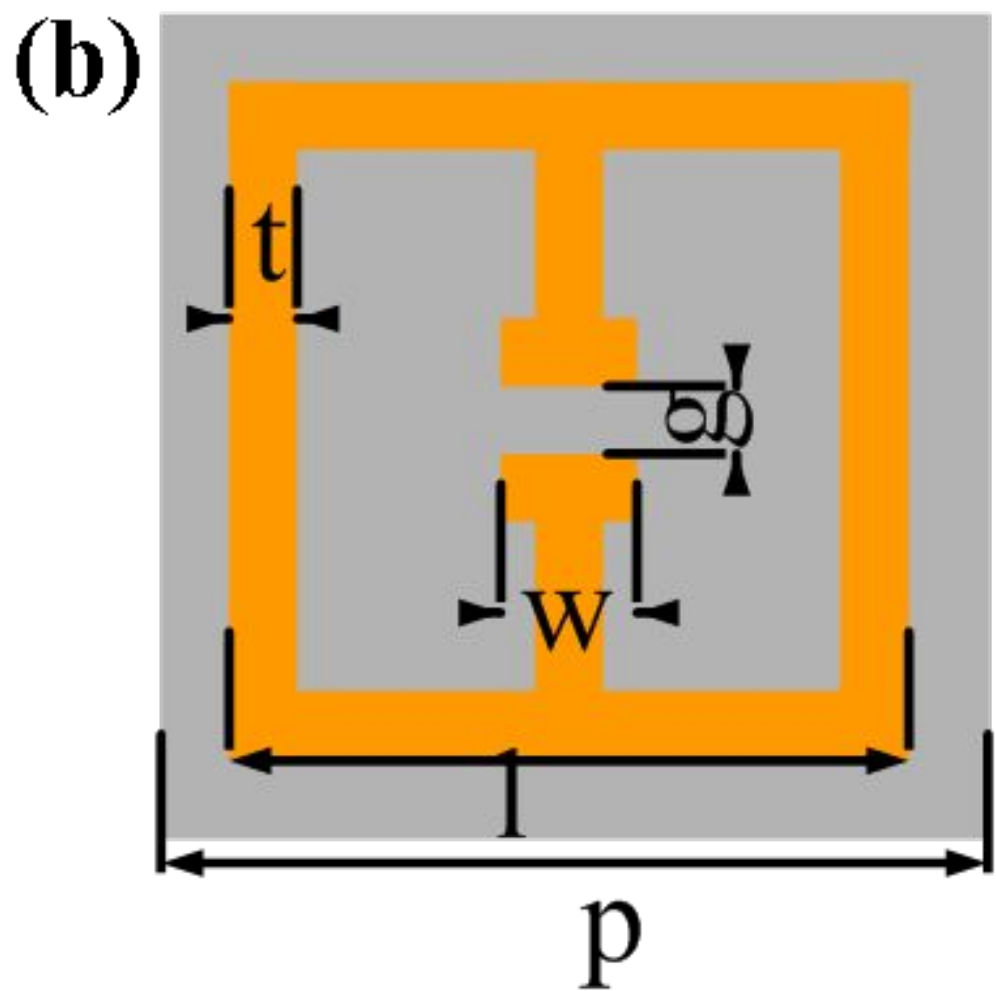

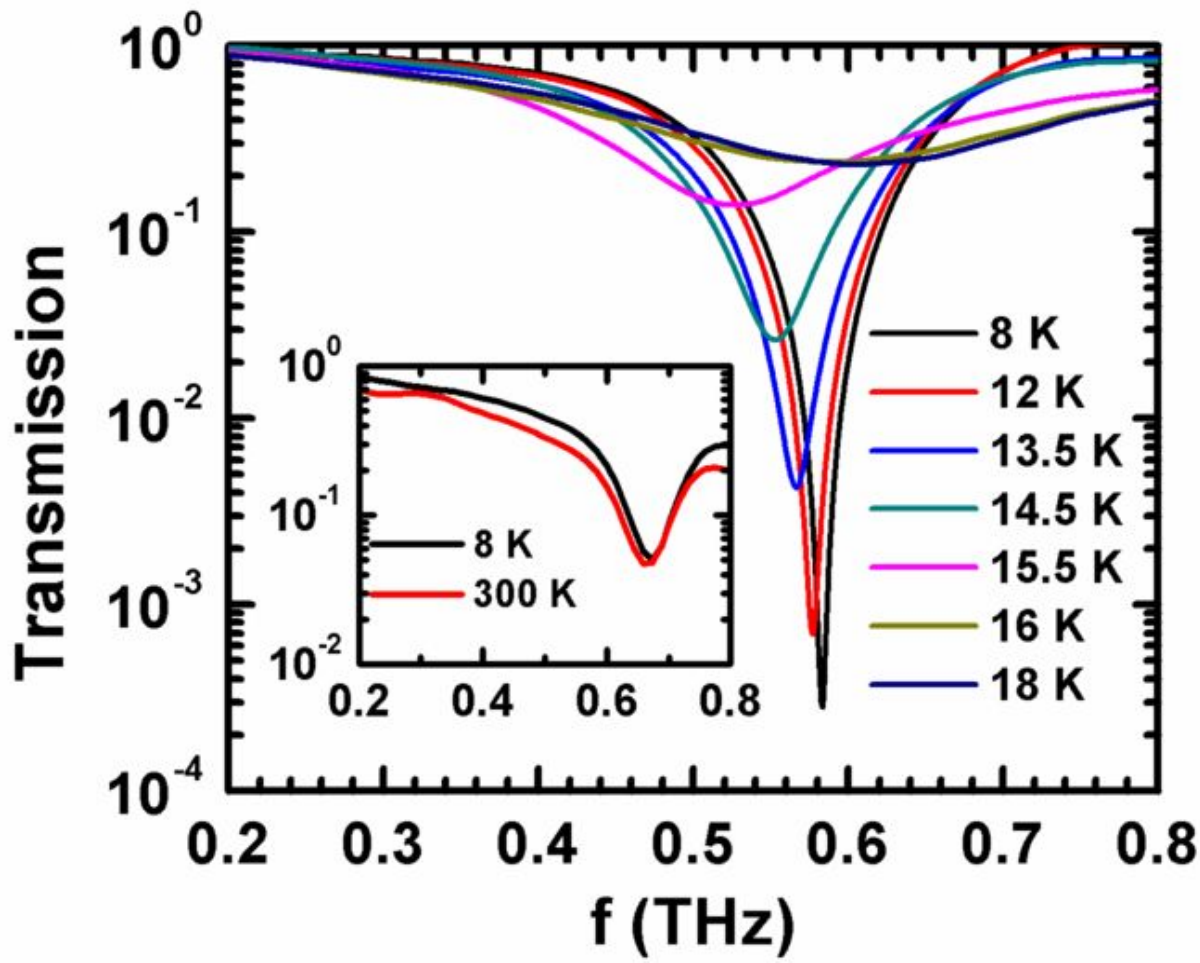

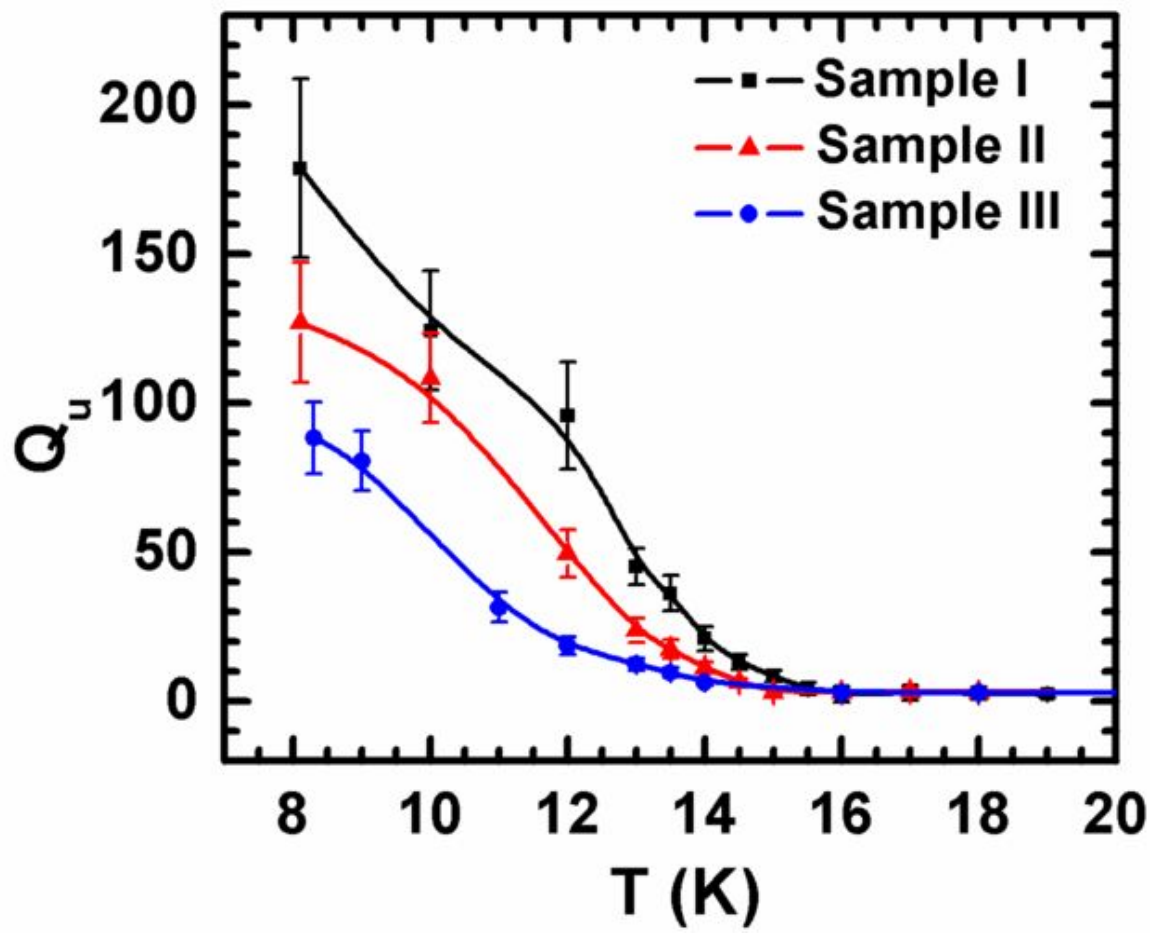

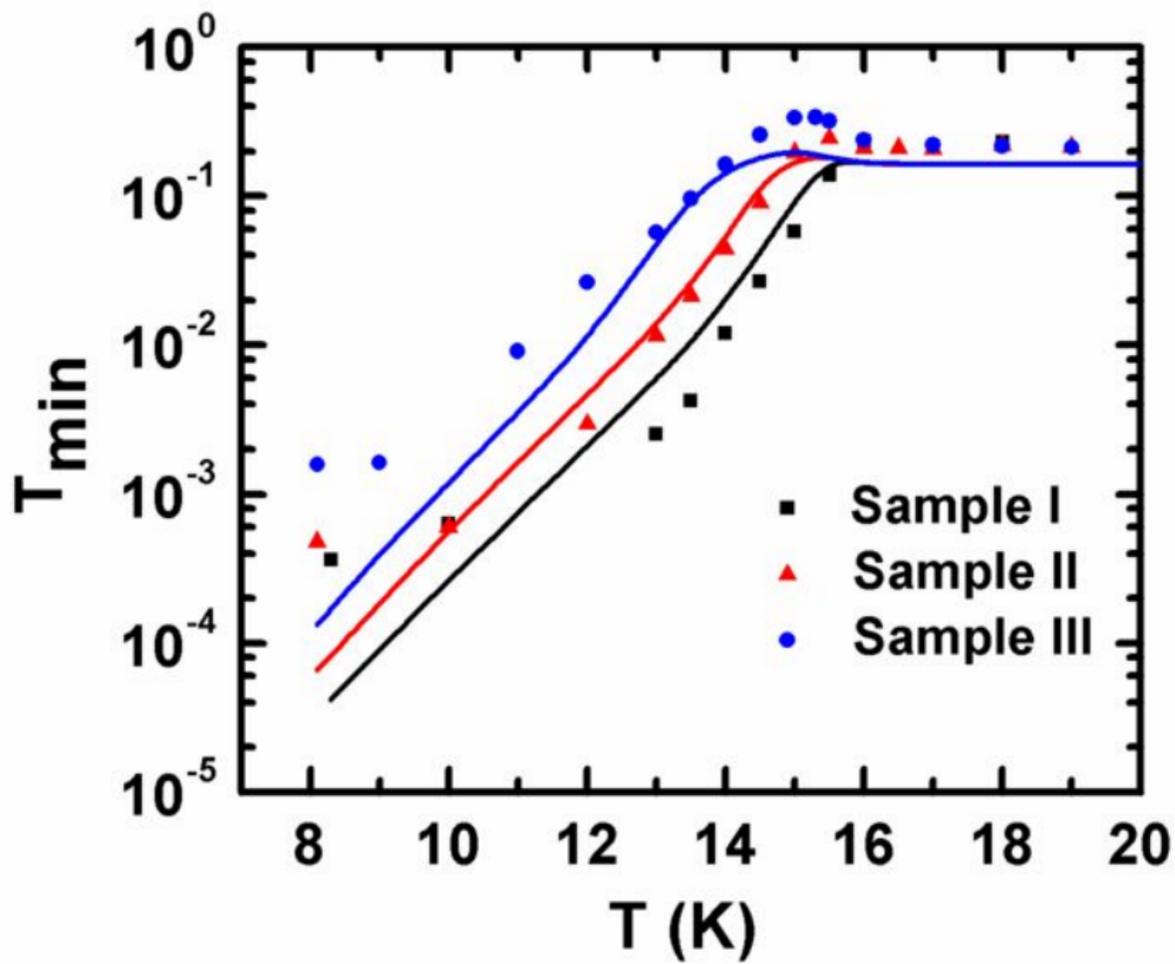